\documentclass[twocolumn,aps,pre,footinbib,floatfix,preprintnumbers,amsmath,amssymb,superscriptaddress,10pt,colorlinks]{revtex4-1}

\usepackage{ulem}
\renewcommand{\emph}{\textit}

\usepackage{selinput}
\usepackage{amsmath}
\usepackage{amssymb}
\usepackage{graphicx}
\usepackage{dcolumn} 
\usepackage{bm} 
\usepackage[usenames,dvipsnames]{color}
\usepackage{natbib}
\usepackage{dsfont}
\usepackage{xr}
\usepackage{mathtools}
\usepackage{here}
\usepackage[urlcolor=RoyalBlue,citecolor=RoyalBlue,linkcolor=BrickRed]{hyperref}

\newcommand{\mean}[1]{\left < #1 \right >}
\newcommand{\abs}[1]{\left | #1 \right |}

\renewcommand{\vec}[1]{\mathbf{ #1 }}

\newcommand{\nabcp}{\underset{\sim}{\nabla}}
\newcommand{\red}[1]{\textcolor{BrickRed}{#1}}

\begin{document}

\title{Mesoscale pattern formation of self-propelled rods with velocity reversal}

\author{Robert Gro{\ss}mann}
\email{grossmann@physik.hu-berlin.de}
\affiliation{Physikalisch-Technische Bundesanstalt, Abbestr. 2-12, D-10587 Berlin, Germany}

\author{Fernando Peruani}
\affiliation{Laboratoire J.~A.~Dieudonn\'{e},~Universit\'{e}~de~Nice~Sophia~Antipolis,~UMR~7351~CNRS,~Parc Valrose,~F-06108~Nice~Cedex~02,~France}

\author{Markus B{\"a}r}
\affiliation{Physikalisch-Technische Bundesanstalt, Abbestr. 2-12, D-10587 Berlin, Germany}

\begin{abstract}
We study self-propelled particles with velocity reversal interacting by uniaxial~(nematic) alignment within a coarse-grained hydrodynamic theory. 
Combining analytical and numerical continuation techniques, we show that the physics of this active system is essentially controlled by the reversal frequency. 
In particular, we find that elongated, high-density, ordered patterns, called bands, emerge via subcritical bifurcations from spatially homogeneous states. 
Our analysis reveals further that the interaction of bands is weakly attractive and, consequently, bands fuse upon collision in analogy with nonequilibrium nucleation processes. 
Moreover, we demonstrate that a renormalized positive line tension can be assigned to stable bands below a critical reversal rate, beyond which they are transversally unstable. 
In addition, we discuss the kinetic roughening of bands as well as their nonlinear dynamics close to the threshold of transversal instability. 
Altogether, the reduction of the multi-particle system onto the dynamics of bands provides a framework to understand the impact of the reversal frequency on the emerging nonequilibrium patterns in self-propelled particle systems.
In this regard, our results constitute a proof-of-principle in favor of the hypothesis in microbiology that reversal of gliding rod-shaped bacteria regulates the occurrence of various self-organized pattens observed during life-cycle phases. 
\end{abstract}

\date{\today}
\pacs{-}
\maketitle

\section{Introduction}

Revealing the physical laws underlying nonequilibrium pattern formation processes in active matter systems, characterized by the permanent conversion of energy into directed motion at the microscale, is central to modern statistical mechanics~\cite{ramaswamy_mechanics_2010,vicsek_collective_2012,romanczuk_active_2012,marchetti_hydrodynamics_2013,menzel_tuned_2015}.  
Dry active matter, composed of self-propelled particles interacting via a velocity alignment mechanism, was classified so far into three potentially different \textit{universality classes}~\cite{ramaswamy_mechanics_2010,marchetti_hydrodynamics_2013}:~polar fluids, self-propelled rods~(SPR) and active nematics~(AN). 
Polar fluids, which have been extensively studied in the context of flocking~\cite{couzin_collective_2002,schaller_polar_2010,deseigne_collective_2010,vicsek_collective_2012,ginelli_intermittent_2015}, are identified by a ferromagnetic alignment symmetry~\cite{vicsek_novel_1995,toner_long_1995,toner_flocks_1998,chate_collective_2008,toner_reanalysis_2012}. 
Systems with nematic~(uniaxial) alignment, namely AN~\cite{ramaswamy_active_2003,chate_simple_2006,ngo_large_2014} and SPR~\cite{peruani_noneq_2006,peruani_mean_2008,baskaran_enhanced_2008,baskaran_hydrodynamics_2008,ginelli_large_2010,peshkov_nonlinear_2012,abkenar_collective_2013,weitz_selfpropelled_2015,nishiguchi_long_2016}, are distinguished by transport properties:~while particles exhibit diffusive back and forth motion at all time-scales in AN, SPR display persistent motion and their instantaneous particle velocity is well-defined. 
We point out that self-propelled particles with nematic velocity alignment that have the ability to reverse their velocity at a finite frequency represent a model system allowing to interpolate between SPR and AN which are contained as limiting cases of low and high reversal frequency, respectively. 
Notably, there are also microbiological systems exhibiting nematic alignment and velocity reversal, e.g.~the bacterial species \textit{Myxococcus xanthus}~\cite{shimkets_induction_1982,starrus_pattern_2012} or \textit{Paenibacillus dendritiformis}~\cite{beer_periodic_2013}. 
In particular, the patterns observed during the lifecycle of myxobacteria depend on the adaption of the reversal rate of individual bacteria~\cite{jelsbak_pattern_2002,igoshin_biochemical_2004,zhang_quantifying_2011}. 
In experiments, non-reversing mutants form large clusters~\cite{peruani_collective_2012}, whereas these large-scale structures break up into a network-like dynamic mesh of one-dimensional nematic streams for the reversing wild-type~\cite{starruss_pattern_2012,thutupalli_directional_2015}.

\begin{figure}[!bp]
  \begin{center}
    \includegraphics[width=0.95\columnwidth]{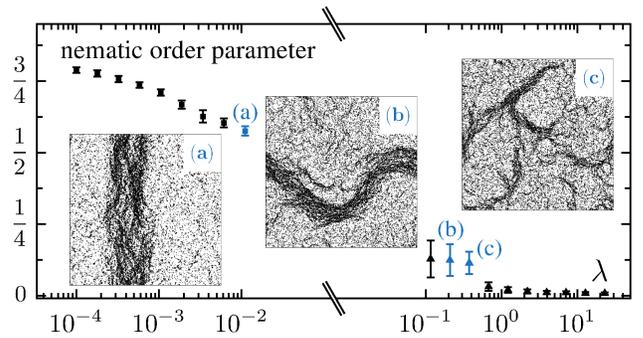}
  \end{center}
   \caption{Nematic order parameter~\cite{chaikin_principles_2000} and corresponding snapshots for simulations of SPR, described by Eq.~\eqref{eqn:model}, for increasing reversal frequency~$\lambda$. The crucial dependence of the degree of orientational order and stability of band structures on reversal is evident. Within high density bands, particles are nematically aligned. Parameters:~$L_{x,y} \!=\! 400$, $\rho_0 \!=\! 0.4$, $N \!=\! 64 \, 000$, $D_{\varphi} \!=\! 0.07$, $\mu \!=\! \pi$, $v_0 \!=\! 1$, $\beta \! \left( \vec{r} \right) \!=\! \Theta \! \left( 1 - \abs{\vec{r}} \right)\!/\pi$, $\Delta t \!=\! 0.01$. } 
  \label{fig:snapshots}
\end{figure}

Since numerical simulations of dry active matter systems revealed the intrinsic link of the emergence of order and formation of large-scale band structures~\cite{chate_collective_2008,ginelli_large_2010,ngo_large_2014,putzig_phase_2014}, we investigate here the mechanisms leading to collective patterns in ensembles of self-propelled particles with nematic alignment and velocity reversal. Our focus is in particular on the influence of reversal frequency on the nematic ordering and bands as depicted in Fig.~\ref{fig:snapshots}. 
A novel analytical expression for periodic band structures is derived and the bifurcation analysis is performed by numerical continuation~\cite{allgower_introduction_2003}. 
Furthermore, it is shown that bands, which do only exist above a critical system size, are rendered transversally unstable for high reversal rates. 
Moreover, we argue that band formation can be understood as a nonequilibrium nucleation process implying attractive band interactions. 
Finally, we discuss the nonlinear stochastic dynamics of bands thereby providing a complete description of the nonequilibrium pattern formation.

\section{Model}

We model~$N$ self-propelled particles with nematic alignment in two dimensions by the Langevin equations 
\begin{subequations}
 \label{eqn:model}
   \begin{align}
    \dot{\vec{r}}_j  &= \vec{v}[\varphi_j] , \label{eqn:model:a} \\
    \dot{\varphi}_j &=\! \mbox{$\sum_{k=1}^N$} \, \mu \beta \! \left ( \vec{r}_{kj} \right) \sin \! \left [2 (\varphi_k - \varphi_j) \right] +\! \sqrt{2D_\varphi}\, \eta_j(t) \label{eqn:model:b}. \!\!
   \end{align}  
\end{subequations}
The velocity of each particle, moving at constant speed~$v_0$, is determined by its direction of motion~$\varphi$ via~$\vec{v}[\varphi]\! =\! v_0 \! \left ( \cos \varphi, \sin \varphi \right )$.
Its position is denoted by~$\vec{r}_j$. 
The interaction may generally depend on the inter-particle distance~$\vec{r}_{kj} \!=\! \vec{r}_k - \vec{r}_j$ reflected by the kernel~$\beta\!\left( \vec{r}_{kj} \right)$. 
We consider short-ranged interactions only:~$\beta \!\left( \vec{r}_{kj} \right)$ vanishes for distances larger than a characteristic length which is rescaled to one without loss of generality. 
Stochastic reorientations of particles, due to spatial heterogeneities for instance~\cite{peruani_self_2007,romanczuk_brownian_2011}, are accounted for by Gaussian fluctuations $\eta_j\!\left( t \right)$ with zero mean, $\mean{\eta_j\!\left (t \right )} \!=\! 0$, and $\delta$-correlations:~$\mean{\eta_j\!\left (t \right ) \! \eta_k\!\left (t' \right )} \!=\! \delta_{jk} \delta \! \left (t-t' \right )$. 
This model is a continuum time version of the Vicsek model with nematic interactions~\cite{ginelli_large_2010}, proposed in~\cite{peruani_mean_2008} as a point-particle model for collective motion of hard SPR~\cite{peruani_noneq_2006}.

As a central element, we additionally include velocity reversals~\cite{igoshin_pattern_2001,boerner_rippling_2002,igoshin_waves_2004,zhang_mechanistic_2012,shi_topological_2013,shi_instabilities_2014,balagam_mechanism_2015,grossmann_diffusion_2016} via $\varphi_j \xrightarrow[]{\; \lambda\;} \varphi_j + \pi$, where~$\lambda$ denotes the reversal frequency.  
We assume a Poissonian reversal process for simplicity, i.e.~stochastic waiting times between two subsequent reversals follow the exponential distribution $\psi(t)\!=\! \lambda e^{-\lambda t}$. 

\section{Hydrodynamic limit}

The large-scale dynamics of self-propelled rods with reversal is addressed within a hydrodynamic theory, which is derived from the Langevin dynamics via the corresponding Fokker-Planck equation~\cite{peruani_mean_2008,farrell_pattern_2012,grossmann_active_2012,*grossmann_vortex_2014}. 
First, we define the coarse-grained one-particle density $p(\vec{r},\varphi,t) \! = \! \mean{\sum_{j} \beta \! \left ( \vec{r} - \vec{r}_j(t) \right ) \! \delta \! \left ( \varphi - \varphi_j(t) \right ) \! }$, where the kernel $\beta\!\left( \vec{r} \right)$ is used for the spatial coarse-graining.
Accordingly,~$p\!\left( \vec{r},\varphi,t \right)$ is slowly varying on scales comparable to the interaction range. 
The Fokker-Planck equation contains two parts, 
\begin{equation}
  \label{eqn:FP_structure}
 \partial_t p(\vec{r},\varphi,t) = \lambda \! \left [ \textcolor{white}{\frac{a}{b}} \!\!\!\! p \! \left (\vec{r},\varphi-\pi,t \right ) - p \! \left (\vec{r},\varphi,t \right ) \right] + \mathcal{L}[p] ,  
\end{equation}
where the first one accounts for reversals and $\mathcal{L}[p]$ is the Fokker-Planck operator for SPR:
\begin{align}
  \label{eqn:FP_rods}
 \mathcal{L}[p] \approx & - \! \mu\, \partial_{\varphi} \! \left [ \int_0^{2\pi} \!\!  d\varphi' \sin \! \left [2 \! \left (\varphi' - \varphi \right ) \right ]  p \! \left (\vec{r},\varphi,t \right ) p \! \left (\vec{r},\varphi'\!,t \right ) \right ] \nonumber \\
 & - \vec{v}[\varphi] \! \cdot \! \nabla p(\vec{r},\varphi,t) + D_\varphi \partial^2_{\varphi} p(\vec{r},\varphi,t) .
\end{align}
The derivation of Eq.~\eqref{eqn:FP_rods} relies on the assumptions that $p\!\left( \vec{r},\varphi,t \right)$ varies slowly in space -- valid by construction -- and that the probability to find two particles at position~$\vec{r}$ with orientations~$\varphi$ and~$\varphi'$ factorizes into the product of one-particle densities in the interaction integral.
This constitutes a mean-field approximation~\cite{dean_langevin_1996,grossmann_superdiffusion_2016}:~we focus on the deterministic part of an actual stochastic field theory~(saddle point approximation~\cite{zinn_quantum_2002}) for the microscopic density $\tilde{p}=\sum_{j=1}^N \beta \! \left( \vec{r} - \vec{r}_j(t) \right) \delta \! \left( \varphi - \varphi_j(t) \right)$. 
The theory can be improved by incorporating noise terms to explain fluctuation-induced shifts of transition points~\cite{solon_revisiting_2013,grossmann_superdiffusion_2016} or the stability of homogeneous, ordered phases in the thermodynamic limit~\cite{kardar_statistical_2007,toner_flocks_1998,toner_reanalysis_2012,ramaswamy_active_2003,chen_critical_2014}. 

Hydrodynamic equations are obtained from the Fokker-Planck equation via a Fourier mode decomposition with respect to the angular variable $\varphi$. 
The Fourier coefficients $f_n\!\left( \vec{r},t \right)\!=\!\int_0^{2\pi}d\varphi \, p\!\left( \vec{r},\varphi,t \right) \! e^{i n \varphi}$ are directly related to local order parameters:~$f_0\!\left( \vec{r},t \right)$ determines the density, $f_1\!\left( \vec{r},t \right)$ corresponds to the polar order parameter and $f_2\!\left( \vec{r},t \right)$ determines the degree of nematic order, accordingly.
Their dynamics is cross-coupled to other modes. 
We reduce this infinite hierarchy to the slow dynamics of the most relevant fields by an appropriate closure relation that allows to express irrelevant fields by the slow variables. 
A closure relation basically entails an assumption about the local properties of a given state~--~it encodes a characteristic lengthscale or, in other words, the closure depends on the smallest lengthscales that a hydrodynamic theory can resolve. 
Since the nematic alignment interaction in Eq.~\eqref{eqn:model} implies that nematic order is predominant on mesoscopic scales and polar clusters are only found on small scales, we identify the particle density~$f_0\!\left( \vec{r},t \right)$ and the nematic order parameter~$f_2\!\left( \vec{r},t \right)$ as relevant fields and eliminate other modes~($\abs{n}\!\neq0,2$) via $\partial_t f_n\!\left( \vec{r},t \right)= 0$ keeping the leading order terms.

It is convenient to work with natural length- and timescales henceforth by rescaling time, length and amplitudes of the fields by $l \! = \! v_0/\!\sqrt{32 D_\varphi\!\left( 2\lambda + 9 D_\varphi \right)}$, $\tau \! = \! 1/(8D_\varphi)$ and $\mathcal{A} \!=\! 8D_\varphi/\mu$, respectively, via $f_0\!\left( \vec{r},t \right)\!=\!\mathcal{A} \rho\!\left( \vec{r}/l,t/\tau \right)$ and $f_2\!\left( \vec{r},t \right)\!=\!\mathcal{A} Q\!\left( \vec{r}/l,t/\tau \right)$.
We eventually obtain the hydrodynamic limit of the microscopic model
\begin{subequations}
  \label{eqn:AN_resc}
  \begin{align}
    \partial_t \rho &\!=\! \mathcal{D} \! \left[ \Delta \rho + \Re \! \left( \nabcp^2 Q^* \right) \right] \label{eqn:AN_rescA}\!, \\
    \partial_t  Q   &\!=\! \frac{\mathcal{D}}{2} \nabcp^2 \!\rho +\! \left( \hspace{-0.006\columnwidth} 1 \!+\! \frac{\mathcal{D}}{2} \right)\!\Delta Q \! + \! \left[ \hspace{-0.000\columnwidth}\left( \hspace{-0.006\columnwidth}\rho - \frac{1}{2} \right) \! - \abs{Q}^2 \right] \! Q ,  \label{eqn:AN_rescB}
  \end{align}
\end{subequations}
where $\nabcp \! = \! \partial_x + i \partial_y$ denotes the Wirtinger derivative~\cite{wirtinger_formalen_1927}. 
The control parameters are the effective density $\bar{\rho}_0 \!=\! \mu \rho_0/\!\left(  8 D_\varphi \right)$, the rescaled system size $\bar{L}_{x,y} \!=\! L_{x,y} \sqrt{32 D_\varphi\!\left( 2\lambda \!+\! 9 D_\varphi \right)} /v_0$ and the coupling coefficient $\mathcal{D} \!=\! \left( 4 \lambda \!+\! 18 D_\varphi \right) \!/\! \left( 2\lambda \!+\! D_\varphi \right)$. 
Accordingly, transport properties are crucially affected by reversal, speed and rotational noise as reflected by~$\bar{L}_{x,y}$:~small~$\lambda$ and high~$v_0$ render the actual system size small.

The closure approximation has another important consequence:~Eq.~\eqref{eqn:AN_resc} has the form of a reaction-diffusion system~\cite{cross_pattern_1993}, in fact it reduces to the field equations for active nematics~\cite{mishra_dynamics_2009}~--~derived previously from the Vicsek model for active nematics~\cite{chate_simple_2006} via the Boltzmann-Ginzburg-Landau approach~\cite{bertin_mesoscopic_2013,ngo_large_2014,peshkov_boltzmann_2014}~--~even though the small scale transport of individual particles is convective~[cf.~Eqs.~\eqref{eqn:model},\eqref{eqn:FP_rods}]. 
This paradox is resolved by noting that particles flip their velocity~--~driven by rotational noise or reversals~--~in a nematic state without affecting the local dynamics which is therefore independent of~$\lambda$. 
Due to velocity reversal, macroscopic transport is diffusive~--~the derived hydrodynamic equations are valid provided that the distance travelled by a particle in between reversals~$v_0/\lambda$ remains considerably smaller than the system size. 

\section{Spatially homogeneous solutions} 
\label{sec:hom_stab_band}

As particle-based simulations suggest, cf.~Fig.~\ref{fig:snapshots}, the collective dynamics is determined by the emergence of large-scale density instabilities. Spatially homogeneous states ($\rho\!\left( \vec{r},t \right)\!=\!\bar{\rho}_0$) both, disordered $Q\!\left( \vec{r},t \right) \!=\! 0$ and ordered $\abs{Q\!\left( \vec{r},t \right)}\!=\!\sqrt{\bar{\rho}_0 - 1/2}$ do not contribute to the understanding of the observed pattern formation phenomena. 
This is a feature shared by several active systems~\cite{boltzmann_bertin_2006,bertin_hydrodynamic_2009,peshkov_nonlinear_2012,grossmann_self-propelled_2013,ngo_large_2014,caussin_emergent_2014,putzig_phase_2014,solon_pattern_2015,ihle_kinetic_2011,ihle_large_2015,ihle_chapman_2016,mishra_dynamics_2009,bertin_mesoscopic_2013}. 
Homogeneously ordered solutions have only been reported for parameter values far away from the order-disorder transition~\cite{toner_long_1995,toner_flocks_1998,boltzmann_bertin_2006,bertin_hydrodynamic_2009,ginelli_large_2010}. 
For the system analyzed here, the disordered homogeneous solution gets destabilized at~$\bar{\rho}_0 = 1/2$. In the vicinity of this point, the homogeneously ordered state is also unstable~\cite{mishra_dynamics_2009,bertin_mesoscopic_2013,putzig_phase_2014} with respect to perturbations that are orthogonal to the orientation of the nematic director~(assumed to be parallel to the~$y$-axis without loss of generality) for
\begin{align*}
  \bar{\rho}_0 \! \in \! \left[ \frac{5}{8} \! - \! \frac{2 \pi^2}{\bar{L}_x^2} \! - \! \frac{\sqrt{\bar{L}_x^2 - 32 \pi^2}}{8\bar{L}_x},\frac{5}{8} \! - \! \frac{2 \pi^2}{\bar{L}_x^2} \! + \! \frac{\sqrt{\bar{L}_x^2 - 32 \pi^2}}{8\bar{L}_x} \, \right] 
\end{align*}
as shown by a blue line in the phase diagram~(Fig.~\ref{fig:phase_diag}).

\section{Emergence of bands}

We analyze now the hydrodynamic theory in one dimension with regard to straight band solutions.
The coordinate system is oriented such that bands are parallel to the $y$ axis~(cf.~Fig.~\ref{fig:snapshots}\red{a}) and, consequently, $\rho\!\left( \vec{r},t \right) \!=\! \rho(x,t)$ and $Q\!\left( \vec{r},t \right) \!=\! Q(x,t)$. 
Since particle-based simulations show that high density and nematic order are intrinsically linked, it is insightful to reduce Eqs.~\eqref{eqn:AN_resc} to the dynamics of the band profile~$B(x,t) \!=\! -Q(x,t)$. 
The minus sign is introduced for convenience such that $B \!\ge\! 0$.
We seek to express the particle density inside a band, denoted by~$\rho \!=\! \rho_B$, by the profile~$B$. 
From Eq.~\eqref{eqn:AN_rescA}, $\rho_B$ is obtained by setting~$\partial_t \rho_B \!=\! 0$, whose solution yields~$\rho_B \! = \! \bar{\rho}_0 \! + \! B \! - \! M[B]$. The band mass $M[B] \!=\! \bar{L}_x^{-1} \!\int_{-\bar{L}_x/2}^{+\bar{L}_x/2} dx' \, B(x'\!,t)$ ensures the global particle number conservation.  
Inserting this ansatz into Eq.~\eqref{eqn:AN_rescB}, the dynamics  
\begin{align}
  \label{eqn:Schloegl}
  \partial_t  B  = \partial_{x}^2 B  +  \left( - \gamma[B] + B - B^2 \right ) \! B 
\end{align}
for the band profile is obtained, where $\gamma[B] \! = \! 1/2 \! - \! \bar{\rho}_0 \! + \! M[B]$, notably independent of $\mathcal{D}$. 
Eq.~\eqref{eqn:Schloegl} is a variant of the Schl\"ogl model, a reaction-diffusion equation with bistable local dynamics. It contains an additional global feedback~\cite{schloegl_chemical_1972,malchow_noise_1985,schimansky_domain_1991,schimansky_analysis_1995} via $\gamma[B]$ ensuring particle number conservation. 
Thus, the hydrodynamic theory in one dimension and the Schl\"ogl model possess the same stationary solutions as well as similar bifurcations points. 
Notably, the dynamics cannot be understood in terms of a free energy minimization thereby underlining the nonequilibrium nature of the temporal dynamics.

\begin{figure}[!tp]
\begin{center}
    \includegraphics[width=\columnwidth]{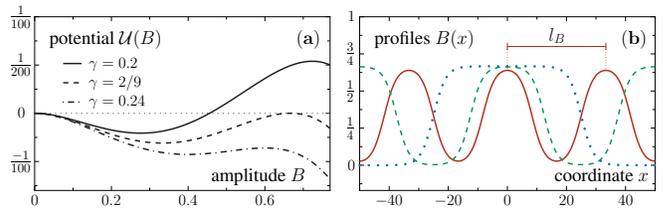}
\end{center}
  \caption{(a)~Illustration of the potential~$\mathcal{U}(B)$ for several values of $\gamma$, which determines stationary band solutions. (b)~Band profiles~$B(x)$ of periodicity~$l_B\!=\!33$~(red solid line), $l_B\!=\!50$~(green dashed line) and $l_B\!=\!100$~(blue dotted line) that coexist at the density $\bar{\rho}_0=11/18$, cf.~Eq.~\eqref{eqn:exact_band11_18}. }
  \label{fig:bif_1d_a}
\end{figure}

\begin{figure}[!tp]
	\begin{center}
		\includegraphics[width=0.95\columnwidth]{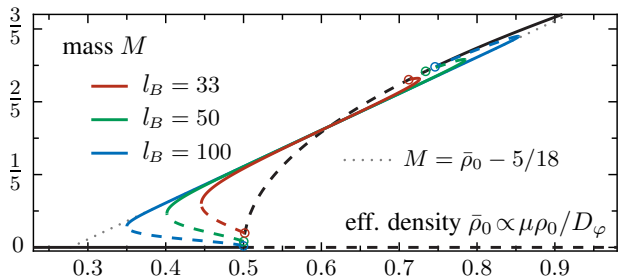}
	\end{center}
	\caption{Bifurcations of band solutions:~stable solutions are shown by solid lines, linearly unstable solutions by dashed lines. Black lines represent the bifurcations of spatially homogeneous states. The mass of stable bands $M\!=\!\bar{\rho}_0-5/18$ as predicted by the Schl\"ogl model is indicated by a grey dotted line. }
	\label{fig:bif_1d_b}
\end{figure}

Setting $\partial_t B = 0$, the problem of finding stationary band solutions is mapped to the motion of a particle in a potential $\mathcal{U}(B) \!=\! - \gamma B^2\!/2 + B^3\!/3 - B^4\!/4$, where $B$ plays the role of position and $x$ is time:~$B''(x) \!=\! -d\,\mathcal{U}(B)/dB$. 
This potential is represented in Fig.~\ref{fig:bif_1d_a}\red{a} for several values of~$\gamma$.
Band solutions are found in analogy to closed orbits in classical mechanics~\cite{goldstein_classical_2002}.
The family of periodic solutions 
\begin{equation}
      \label{eqn:exact_band11_18}
      B(x;m) = \frac{1}{3} \left[ 1 + \sqrt{\frac{2m}{m+1}} \,\mbox{cd} \! \left( \frac{x}{3\sqrt{m+1}},m \right) \right] \! ,  
\end{equation}
parametrized by $m \! \in \! (0,1)$, is found analytically for $\gamma \! = \! 2/9$, corresponding to the global density $\bar{\rho}_0 \!=\! 11/18$.
The periodicity of these bands~(Fig.~\ref{fig:bif_1d_a}\red{b}) is determined by~$l_B(m)\!=\!12 \sqrt{m+1}\!\cdot\!F\!\left(\pi/2,m\right)$, where~$F(x,m)$ denotes the elliptic integral of the first kind and~$\mbox{cd}\!\left(x,m\right)$ is a Jacobi elliptic function~\cite{NoteFoot1}. 
Besides this one-parametric family of periodic solutions, a homoclinic solutions exists~($l_B\!\rightarrow\!\infty$)~--~relevant in the thermodynamic limit~--~which was studied in~\cite{bertin_mesoscopic_2013}. 
A band with periodicity~$l_B$ can exist in a system size of length~$\bar{L}_x$ if the latter is an integer multiple of~$l_B$. 
We note that bands, represented by oscillations around the minimum of~$\mathcal{U}(B)$, emerge above a critical system size only since the minimal period of oscillations -- corresponding to~$l_B$ -- is nonzero for harmonic oscillations.  
The numerical continuation~\cite{allgower_introduction_2003} of the analytical solutions reveals that bands emerge via two subcritical bifurcations~(Fig.~\ref{fig:bif_1d_b}).
Increasing~$\bar{\rho}_0$, we observe a~(i)~linearly stable disordered state, (ii) disordered state coexisting with band solutions, (iii)~family of band solutions, (iv)~bands coexisting with the ordered state and (v)~a linearly stable ordered state. 
This is also summarized in the phase diagram, see~Fig.~\ref{fig:phase_diag}, in accordance with direct simulations of the hydrodynamic equations in~\cite{putzig_phase_2014}.

So for parameters for which bands of different period coexist on the deterministic level, the question arises which of these states is most likely observed in a particle-based Langevin simulations including noise. 
Since the hydrodynamic theory can be mapped to the Schl\"ogl model~[Eq.~\eqref{eqn:Schloegl}], which is a generic model for Ostwald ripening~\cite{schimansky_domain_1991}, we expect bands to merge when they come close to each other.  
Hence, band interaction is attractive. 
Indeed, the situation shown in Fig.~\ref{fig:snapshots} is common, i.e.~one band develops in the course of time~\cite{ginelli_large_2010} for intermediate system sizes. 
However, wide bands possess exponential tails such that the interaction of well separated bands is weak. 
The creation and fusion of bands is driven by noise in this regime. 

\section{Transversal band dynamics}

\begin{figure}[!tp]
    \begin{center}
      \includegraphics[width=\columnwidth]{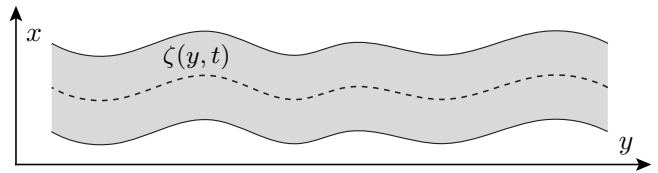}
    \end{center}
    \caption{Illustration of a transversally modulated band. The dashed line indicates the filament~$\zeta\!\left( y,t \right)$. }
    \label{fig:band_2d}
  \end{figure}

How does a band dynamically evolve given that an initially straight band is weakly modulated transversally? 
Several responses are conceivable:~a restoring force restabilizes the straight band or fluctuations increase in time. 
We study the transversal band dynamics in terms of the filament~$\zeta(y,t)$ determining the center of the band in every cross section parallel to the $x$-axis~(Fig.~\ref{fig:band_2d}).
Using the band profile solutions, we formulate the following ansatz which is based on the fact that the filament dynamics is slow compared to amplitude fluctuations of the band~(cf.~transversal instabilities of reaction-diffusion fronts~\cite{kuramoto_chemical_2012,Malevanets_biscale_1995}): 
\begin{subequations}
  \label{eqn:2d:ansatz}
  \begin{align}
    \rho(\vec{r},t) &\simeq \rho_B\! \left (x - \zeta(y,t) \right ) + \delta \rho(\vec{r},t), \\
    Q(\vec{r},t)    &\simeq -B\! \left (x - \zeta(y,t) \right ) + \delta Q(\vec{r},t) .
  \end{align} 
\end{subequations}
The correction terms~$\delta \rho$ and~$\delta Q$ account for deformations of band profiles which appear once the band is curved. 
Hence, these corrections must equal zero for straight bands~($\zeta \! = \! \mbox{const.}$). 
Accordingly, we expand the perturbations in small gradients of~$\zeta$ as  
\begin{subequations}
  \label{eqn:2d:expansion}
  \begin{align}
    \delta \rho(\vec{r},t) &\simeq \mbox{$\sum_{k=1}^\infty$} \delta \rho_k(x) \partial^k_y \zeta(y,t), \\
    \delta Q(\vec{r},t)    &\simeq \mbox{$\sum_{k=1}^\infty$} \delta Q_k(x)    \partial^k_y \zeta(y,t).     
  \end{align} 
\end{subequations}
Along similar lines, the linear filament dynamics is written:
\begin{equation}
  \label{eqn:fil_lin}
  \partial_t \zeta\!\left( y,t \right) \simeq - \mbox{$\sum_{k=1}^\infty$} K_{2k} \partial^{2k}_y \zeta(y,t) . 
\end{equation}
By inserting Eqs.~\eqref{eqn:2d:ansatz}-\eqref{eqn:fil_lin} in the hydrodynamic theory~[Eq.~\eqref{eqn:AN_resc}] and collecting terms of similar order in~$\partial_y^k \zeta$, the functions $\delta \rho_k$, $\delta Q_k$ as well as the coefficients $K_{k}$ are perturbatively accessible enabling the construction of the two-dimensional band solution.  

\begin{figure}[!tp]
  \begin{center}
    \includegraphics[width=\columnwidth]{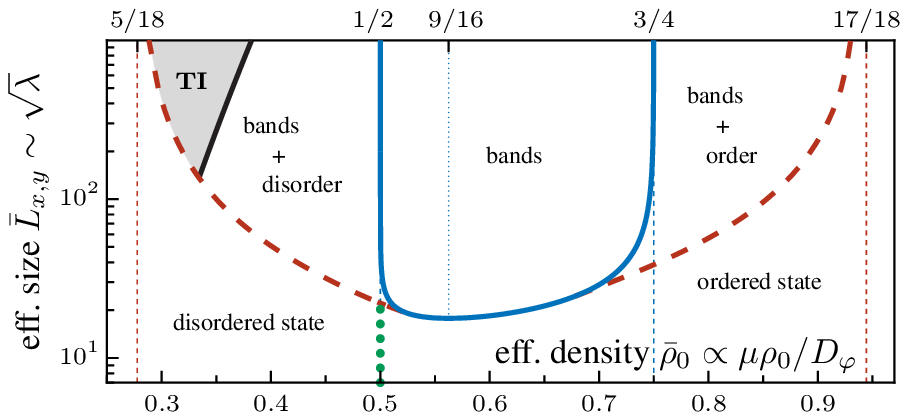}
  \end{center}
  \caption{Phase diagram as a function of~$\bar{\rho}_0\!\propto\!\rho_0\mu/D_\varphi$ and the effective system size~$\bar{L}_{x,y}\!=\!L_{x,y}\sqrt{32D_\varphi\!\left(2\lambda\!+\!9 D_\varphi\right)}/v_0$ for~$\mathcal{D} \!=\! 10$. Thick lines correspond to bifurcation points:~the mean-field order-disorder transition for~$\bar{\rho}_0 \!=\! 1/2$ is shown in green~(dotted); the blue line~(solid) indicates the transversal instability of the homogeneously ordered state~(derivation not shown, cf.~\cite{mishra_dynamics_2009,bertin_mesoscopic_2013,putzig_phase_2014}) and section~\ref{sec:hom_stab_band}; the red~(dashed) line shows the saddle-node bifurcation of bands as shown in Fig.~\ref{fig:bif_1d_b} obtained by numerical continuation of the analytical band solution~[Eq.~\eqref{eqn:exact_band11_18}]. The parameter region corresponding to transversally unstable bands is shown in grey~(TI) as predicted from the linearized filament dynamics~[Eq.~\eqref{eqn:fil_lin}]. Since bands are always transversally unstable for~$\bar{L}_{x,y} \! \rightarrow \infty$~\cite{ngo_large_2014}, the black solid line will span over the whole range of band existence (within the red, dashed lines). }
  \label{fig:phase_diag}
\end{figure}

Now, we restrict the analysis to the second linear order.
The linear stability of a straight band~$\zeta \! = \! \zeta_0$ is determined by the dispersion relation $\sigma(q_n) \!\simeq \! K_2 q_n^2 - K_4 q_n^4$, constituting the exponential growth rate of a mode corresponding to the wavenumber $q_n \!=\! 2\pi n/\bar{L}_y$ with~$n \!\in\! \mathds{Z}$.
Due to the translational symmetry, the $q_0$-mode -- corresponding to translational shifts -- is neutral meaning $\sigma\!\left( q_0 \right) \!=\! 0$ implying that~$\zeta(y,t)$ is a slow variable.
Our analysis reveals that both,~$K_2$ and~$K_4$, are positive, thus suggesting that modes in the range $q_n^2 \!\in\! ( 0, K_2/K_4 )$ are unstable.
Accordingly, an instability occurs for $\bar{L}_y\! >\! 2 \pi \sqrt{K_4/K_2}$. 
This includes the transversal instability in the thermodynamic limit as suggested in~\cite{ngo_large_2014}. 
The estimate for $\bar{L}_y$ allows us to  complete the phase diagram by indicating the region of transversally unstable bands~(grey-shaded region in Fig.~\ref{fig:phase_diag}). 
Notice that bands are stable in a wide range of parameter space thus explaining why stable bands are found in particle-based simulations.  

Finally, we comment on the stochastic, nonlinear dynamics of bands. 
The lowest order nonlinearity which is compatible with all symmetries reads $\partial_y (\partial_y \zeta)^3$. 
Note that the KPZ-like nonlinearity $\left( \partial_y \zeta \right)^2$ is ruled out here by the mirror symmetry
$\zeta\to-\zeta$~\cite{kardar_dynamic_1986,barabasi_fractal_1995}. 
Below the transversal instability, all modes $q_n$ are larger than $q_c \! = \! \sqrt{K_2/K_4}$, enabling the approximation~$\sigma(q_n) \! \approx \! - K_4 q_n^4$. 
Hence, the stochastic nonlinear band dynamics below the transversal instability reads 
\begin{equation}
  \partial_t \zeta \simeq - K_4 \partial_y^4 \zeta + \chi \partial_y \!\left ( \partial_y \zeta \right )^3 + \sqrt{2D} \, \xi(y,t) ,
\end{equation}
where~$\chi$ is a phenomenological parameter. In the context of surface growth and molecular beam epitaxy~\cite{sarma_solid_1992,*Csarma_solid_1992,*Rsarma_solid_1992,krug_origins_1997}, it has been argued~\cite{lai_kinetic_1991,sarma_dynamical_1994,kshirsagar_nonlinearities_1996,kim_dynamical_1995} that this equation describes roughening according to the Edwards-Wilkinson~(EW) class~\cite{edwards_surface_1982}. Thus, stable bands exhibit an effective positive line tension $\tilde{\nu}$ facilitating to recast the filament dynamics in the form~$\partial_t \zeta \simeq \tilde{\nu} \partial_y^2 \zeta \!+\! \sqrt{2\epsilon} \, \xi$. 

In order to address the dynamics close to the threshold of transversal instability, it is insightful to introduce the field $\Lambda(y,t) = \partial_y \zeta(y,t)$ obeying 
\begin{equation}
  \label{eqn:modB}
  \partial_t \Lambda(y,t) \simeq \partial^2_{y} \! \left[ \chi \Lambda^3 - K_2 \Lambda - K_4 \partial^2_y \Lambda \right]\! + \eta_c(y,t),
\end{equation}
where $\eta_c(y,t)$ denotes a conserved white Gaussian noise~\cite{tauber_critical_2014}. 
Accordingly, the dynamics of~$\Lambda(y,t)$ is determined by the stochastic \textit{model} $B$~\cite{hohenberg_theory_1977}, also known as \textit{Cahn-Hilliard} equation~\cite{cahn_free_1958}.
Thus, the derivative of~$\zeta$ is determined locally by~$\Lambda \!=\! \pm \sqrt{K_2/\chi}$ for positive~$\chi$. 
The dynamics described by Eq.~\eqref{eqn:modB} implies the attraction of points with similar signs of~$\Lambda$ leading to a piece-wise constant derivative and, hence, a zig-zag shaped band with at least two turning points~(cf.~Fig.~\ref{fig:snapshots}\textcolor{BrickRed}{b}). 
Eventually, bands are most likely to break apart at these turning points, where the curvature is maximal. 
Transversally unstable bands may restabilize in rectangular domains along the shortest dimension of the system. 

\section{Discussion \& outlook}

We studied self-propelled rods with velocity reversal, whereby we focused in particular on the emergence and dynamics of nematic band structures. 
A central step of the analysis was the reduction of the corresponding hydrodynamic field equations to a modified Schl\"ogl model with global feedback. 
We note that, overall, the phase separation process with the associated symmetry breaking are generic nonequilibrium phenomena without analogues in equilibrium statistical mechanics. 

The pattern formation approach adopted in this study is suitable to address emergence and stability of structures in active systems of finite size.
It is important to stress that the analysis of  the  existence of homogeneous ordered phases in the thermodynamic limit requires a field theoretic analysis including fluctuations, enabling to understand how information travels in the system~\cite{toner_long_1995,toner_flocks_1998,toner_reanalysis_2012,ramaswamy_active_2003,grossmann_superdiffusion_2016}.  
Whereas the large-scale transport is,  for finite reversal rates and $v_0/\lambda\ll L$, arguably diffusive, implying quasi long-range order~\cite{ramaswamy_active_2003}, transport properties in the thermodynamic limit for vanishing reversal and the related uniqueness of the SPR universality class remain open theoretical challenges, as recent experiments suggest the emergence of long-range order~\cite{nishiguchi_long_2016}.

Concerning experiments, the finding that reversing self-propelled rods self-segregate into elongated nematic streams suggests a potential mechanism for pattern formation in microbiological systems such as myxobacteria:~since unstable bands may be rendered stable for low reversal frequencies~(see Fig.~\ref{fig:snapshots}), the initial stage of aggregation could be triggered at the individual level by downregulating the reversal frequency, in line with experimental studies~\cite{jelsbak_pattern_2002} and corresponding simulations~\cite{ginelli_large_2010,balagam_mechanism_2015}. 
Experiments further revealed an increasing mean particle speed during the aggregation~\cite{jelsbak_pattern_2002}, in turn restabilizing large-scale structures according to our theory.
Therefore, the present study suggests that the regulation of velocity reversal is a key element to understand aggregation of several microbiological species. 
The detailed modeling of these systems may require more realistic models, in particular the consideration of hydrodynamic interactions, the exchange of chemical signals, heterogeneous environments or boundary effects, thereby offering a plethora of potential extensions of the present study.   

\begin{acknowledgments}

We thank Lutz Schimansky-Geier, Harald Engel and Igor Sokolov for valuable discussions and critical remarks. R.G.~and M.B.~acknowledge the support by the German Research Foundation via Grant No.~GRK~1558. F.P.~acknowledges support from Agence Nationale de la Recherche via Grant ANR-15-CE30-0002-01. 

\end{acknowledgments}

%


\begin{thebibliography}{92}
\bibitem [{Not()}]{NoteFoot1}%
  \BibitemOpen
  \href@noop {} {}\bibinfo {note} {{\textnormal{We use the convention $F(x,m)
  \!=\!\int_0^x d\theta \left (\! \sqrt{ 1-m \sin^2 \theta } \,\right )^{-1}$
  for the elliptic integral, whose inverse function -- the Jacobi amplitude --
  is abbreviated by $\mbox{am}\! \left( x,m \right)$. The Jacobi elliptic
  function $\mbox{cd}\!\left( x,m \right)$ is determined by the Jacobi
  amplitude via $\mbox{cd}\!\left( x,m \right) = \cos \! \left[ \mbox{am}\!
  \left( x,m \right) \right]\!/\!\sqrt{1-m \sin^2 \! \left[ \mbox{am}\! \left(
  x,m \right) \right]}$.}}}\BibitemShut {Stop}%
\end{thebibliography}
\end{document}